\documentclass{article}
\usepackage{graphicx}

\title{Agency under indefinite causality: operational eternalism in higher-order quantum theory}
\author{Alexei Grinbaum\footnote{CEA-Saclay, 91191 Gif-sur-Yvette Cedex, France}}

\date{}

\begin{document}

\maketitle
\begin{abstract}
After two decades of research on indefinite causality, a conceptual lesson emerges: 
the tension between operational quantum theory and dynamical spacetime physics requires observers, or agents, to be defined within the theory itself. Agents arise from admissible groupings of input and output data in a global Hilbert space, and redefining these groupings may eliminate non-causality. Agency is perspectival: from Alice's perspective, Bob may fail to qualify as an observer, and vice versa. We interpret these results through an \emph{operational eternalism}, a stance analogous to the block-universe view but applied to information rather than geometry. Observers are thus well-defined in the global eternalist picture, but they are not universal across local perspectives.
\end{abstract}

\section{Introduction}

Historically, physical theories have been developed as dynamical theories describing the time evolution of a physical system. The operational approach offers an alternative: an operational theory is not an attempt to predict the future state of a physical system via an evolution equation in continuous time. Rather, it employs probabilities to establish a lawlike dependence between  inputs and outputs obtained by, or assigned to, agents or observers (treated as synonyms in this work). This lawful probabilistic relation provides explanatory power that is not grounded in the knowledge of the dynamics. We propose that indefinite causality between agents in the operational approach compels us toward operational eternalism. Analogous to the block-universe view, operational eternalism treats data as primary givens rather than dynamical becomings, provides a global perspective on observerhood, and renders agency perspectival.

The operational approach need not postulate systems as fundamental entities.
One example is the device-independent setting where systems are not theoretical givens~\cite{grin_devindep}. Instead, an operational theory begins with a distinction between two types of information or data, called inputs and outputs. Historically, `inputs' and `outputs' entered physical jargon as information-theoretic reformulations of parameter settings and measurement outcomes~\cite{Shimony1984} and have since become primitive concepts in operational approaches to quantum theory. Their lawlike relations are often represented diagrammatically:
inputs and outputs are connected by lines or wires that correspond to transformations between ``boxes'', each parameter of the wires representing an input or an output assigned to a particular agent or observer. Systems are not entities of the world, as in a realist account, but a name of ``what propagates'' along such lines --- a conception that leads to ``new modes of explaining physical phenomena''~\cite{coecke_pict}.

\begin{figure}
    \centering
    \includegraphics[width=0.8\linewidth]{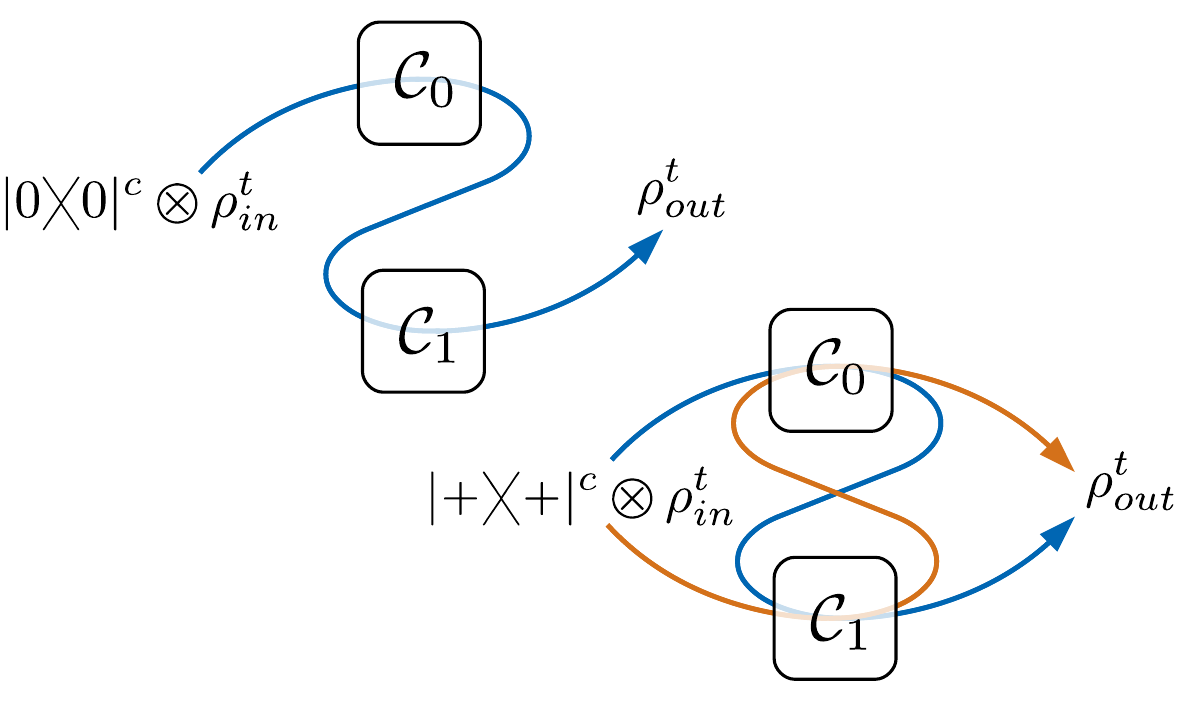}
\caption{
The quantum switch places the order of two operations, $C_0$ and $C_1$, under coherent quantum control \cite{chiribella_quantum_2013,LoizeauAG}. If the control is in the state $| 0 \rangle$, the target undergoes $C_0$ followed by $C_1$. If it is in $|1\rangle$, the order is reversed. Preparing the control in a superposition produces a coherent superposition of the two orders. Conventionally, $C_0$ and $C_1$
are associated with Alice’s and Bob’s laboratories.}
    \label{fig:switch}
\end{figure}

Inspired in part by Lucien Hardy's work on operational foundations of quantum gravity, the research on indefinite causality~\cite{Brukner_rev} began with the invention of an operational setting called a ``quantum switch'' (Figure~\ref{fig:switch}). In this process, the order of passage through Alice's and Bob's lab is controlled by a quantum system. When the control qubit is in a superposition state, this order is indefinite~\cite{chiribella_quantum_2013}. Over time, multiple paradigms for indefinite causality were introduced, including process matrices~\cite{oreshkov_quantum_2012}, supermaps~\cite{Chiribella_2008}, or higher-order quantum theory~\cite{taranto2025higherorderquantumoperations}.

The foundational significance of non-causality is somewhat similar to the impact of non-locality research between the 1960s and the 1990s. Until the Einstein-Podolsky-Rosen paper~\cite{EPR}, locality in space was seen as a necessary condition of physical theory. Later, due to Bell~\cite{Bell1}, it became a central quantifiable indicator partitioning quantum theory from hidden-variable models. With the invention of Popescu-Rohrlich boxes thirty years after Bell~\cite{popescu}, the amount of non-locality came to be seen as a parameter characterizing general probabilistic theories~\cite{hardy,barrett_information_2007,Chiri} to capture their degree of non-classicality~\cite{popescu2014}. As a result of this algebraic reformulation of locality, our understanding of the role of space in quantum theory has significantly improved.
Research on indefinite causality achieves a similar improvement in our understanding of the relation between observers and spacetime due to violation of causal inequalities. 

We draw a foundational lesson from this research: the tension between dynamical theory in spacetime and operational theory is impossible to bridge without bringing agents into the theory and defining them in a perspectival way. The key stumbling block is the impossibility to reconcile operational primitives with the primitive conception of events in spacetime physics. 
This tension implies that taking one agent's perspective may lead to the disqualification of another one as an agent. While Alice and Bob are both agents from the eternalist viewpoint, whether one may still be considered as an agent depends on the other's perspective: from Alice's perspective, Bob may not qualify as an agent, and vice versa.

\begin{figure}[h]
    \centering
\includegraphics[width=\linewidth]{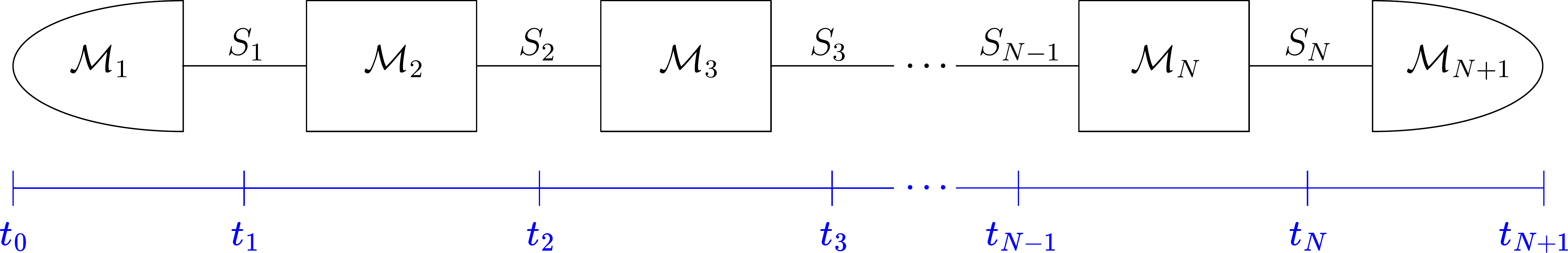}
\caption{
A circuit diagram represents a sequential process relative to an agent’s local temporal perspective \cite{wechs2024}. At each step, a transformation $M_i$
maps the system $S_{i-1}$ at time $t_{i-1}$ to $S_{i}$ at the time $t_i$.
Such a representation presupposes both a decomposition into systems and an operational foliation into temporally ordered transformations. Operational eternalism admits multiple such perspectives without privileging any one of them.}
    \label{fig:circuit}
\end{figure}

It is now possible to formulate precise criteria for one observer to be consistently viewed as an agent from another observer's perspective. An agent's local perspective involves two ingredients: a decomposition into subsystems to place inputs and outputs in respective subspaces, and a foliation into circuit fragments that are sequential in the agent’s local time (Figure~\ref{fig:circuit}). 
Multiple agents may be constituted within the global standpoint of operational eternalism, but their agency need not persist across all local perspectives. Agents are well-defined, but they are not universal.

The emergence of the information-theoretic operational approach came in stages in the history of physics. We recall three such stages in Section~\ref{section:hist}. In Section~\ref{sect3} we briefly summarize key results on indefinite causality. In Section~\ref{sect:eternalism} we advance operational eternalism as an interpretative proposal for higher-order theory, before concluding in 
Section~\ref{sect:conclusion}.

\section{A brief history of operationalism}\label{section:hist}

Accepting the operational approach as a valid form of physical theory was neither immediate nor free of controversy. Epistemological debates in 20th-century physics often focused on what constitutes a theory. This began with Einstein's distinction between constructive and principle theories~\cite{ein19} and continued all the way to the information-theoretic interpretations of quantum theory~\cite{grinbaumInfo}. Quantum theory removed the dogma of analyzing the evolution of a physical system by specifying a set of initial coordinates as positions and momenta in space. The Hilbert space, introduced in quantum theory by von Neumann~\cite{HvNN,vN32}, has paved the way to alternative mathematical frameworks. The epistemological question of deciding which approaches count as valid has loomed large ever since.

We recall three episodes from the history of operational approaches in the foundations of quantum mechanics. All three help emphasize important points that will later resurface in the interpretational debate about indefinite causality, reveal a progressive stripping away of ontological assumptions, and exemplify the reluctance to admit that a physical theory may be built following the operational approach. We begin with a story of a misunderstanding between Einstein and Heisenberg, proceed with a disagreement between Bell and Piron, and conclude with a lesson from Lucien Hardy. Hardy himself enlists Einstein and Heisenberg as early adopters of an operational approach: ``Heisenberg was, of course, very much influenced by the operationalism of Einstein''~\cite{hardy_towards_2007}. The real situation was, however, slightly more complicated.


\subsection{Early operationalism: Einstein vs Heisenberg}


Einstein's `operationalism' was inferred by the young Heisenberg from the senior scientist's insistence on using observed events to construct the theory of relativity: 
\begin{quote}
    ``All our spacetime verifications invariably amount to a determination of spacetime coincidences. Moreover, the results of our measurements are nothing but verifications of such meetings of the material points of our measuring instruments with other material points \ldots and \textit{observed point-events} happening at the same place and the same time. The introduction of a system of reference serves no other purpose than to facilitate the description of the totality of such coincidences''~\cite[our emphasis]{Ein16}.
\end{quote} This operationalism was the best-known example of what Einstein called a ``principle theory'': the principle of relativity narrows the possibilities and acts as a constraint, but it cannot be logically derived from empirical observations~\cite{ein19,EinSolovine}.

Despite Einstein's apparent influence, the idea that quantum theory should be based on observable quantities is unquestionably Heisenberg's. The latter famously argued in 1925 that quantum mechanics should be ``founded exclusively upon relationships between quantities which in principle are observable"~\cite{Heis25}. As Heisenberg recalled in a later memoir, he was certain that Einstein's theory relied on the same assumption. To his  surprise, Einstein objected:
\begin{quote}
    
\textemdash But you don't seriously believe that none but observable magnitudes must go into a physical theory?

\textemdash Isn't that precisely what you have done with relativity?

\textemdash Possibly I did use this kind of reasoning but it is nonsense all the same. On principle, it is quite wrong to try founding a theory on observable magnitudes alone. \ldots It is the theory which decides what we can observe.~\cite[p.~59]{Heis1969}
\end{quote}

Naively, Heisenberg's insistence on using observable quantities in the construction of quantum mechanics could be read as a call to use only empirically given data in the dynamical equations of motion. Einstein's objection was to insist that the search for empirical data should not be based on previous physics but be guided by the new emerging theory. Neither scholar ever questioned the dogma that building a theory meant providing dynamical equations of motion.

Heisenberg's vision did not materialize in the mathematical formalism he created, so he took care to insert ``in principle'' in his memoir. The actual tendency after 1925 was to use a mathematical formalism that seemed increasingly abstract rather than empirical, especially for the proponents of old-style physics. Lorentz objected at the Solvay congress in 1927: \begin{quote}``I am told that by all these considerations one has come to construct matrices that represent what one can observe in the atom, for instance the frequencies of the emitted radiation. Nevertheless, the fact that the coordinates, the potential energy, and so on, are now represented by matrices indicates that the quantities have lost their original meaning and that one has made a huge step in the direction of abstraction''~\cite[pp.~402-403]{Baccia}.
\end{quote}This was a far cry from Heisenberg's supposed empiricism. While the younger physicists, viz. Born and Dirac, were quick to dismiss Lorentz's doubts, Bohr recognized their significance: ``The issue of meaning raised by Mr Lorentz is of very great importance''~\cite{Baccia}. This controversy set the stage for a long debate about the appropriate mathematical formalism of quantum theory~\cite{unreas,grineffectiveness}. From the very beginning, this debate was  focused on the interplay between abstract mathematical tools that are necessary to build the theory and empirically observable quantities. Heisenberg's operationalism will resonate again in the information-theoretic operational approach with its attempt to use mathematics in order to establish a lawful physical connection between data points.

\subsection{Local subsystems: Piron vs Bell}

The interpretations of indefinite causality are not unlike the interpretations of non-locality after John Bell. Ever since Bell's 1964 article~\cite{Bell1}, the meaning of non-locality has been a topic of lively debate in the foundations of physics. Up until the 1980s the interpretations focused on the demise of one of the two fundamental assumptions: parameter independence or outcome independence. Later, the debate shifted toward generalized contextuality and the role of subsystem composition in generalized models featuring different amounts of non-locality. 

John Bell opened his article with a reference to the definition of ``element of reality'' in the work of Einstein, Podolsky, and Rosen~\cite{EPR}. This definition included the idea of ``not disturbing a [distant] system in any way''. Bell interpreted the term ``distant'' as an intent ``to restore causality and locality'' in quantum theory, in agreement with Einstein's own comments. In principle, locality and causality are not synonymous, with the latter responding to Einstein's desire to restore determinism in physical theory. In Bell's later thinking, however, these two principles quickly merged into one, limiting the analysis of causality to the study of local causes. This is what Bell subsequently termed ``locality''~\cite{Bell2}, ``local causality''~\cite{Bell75,Bell76}, or ``local realism''~\cite{Bell1976A}, using the terminology of Clauser, Horne, Shimony, and Holt~\cite{CHSH,CH74}. 

Why was causality dropped as a standalone principle? One likely reason is the connotation that this term had taken in the hidden variable approach: since the early 1950s, David Bohm's theory was known as the ``causal interpretation'' of quantum mechanics~\cite{Bohm1951,Bohm1957}. 
Bell referred to Bohm's work in 1964 without particular admiration; when he occasionally used the term ``causality'' later, he equated it with ``no action at a distance''~\cite{Bell4}. The question of the interplay between local and global causality has never been raised: Bell's focus on local realism was so strong that his initial introduction of causality as a principle separate from locality vanished from the foundational debate for several decades~\cite{LAUDISA202344}.

As Bell's result was gaining notoriety, it met strong objections from the Geneva school of quantum logic, particularly from its new leader, the mathematical physicist Constantin Piron, who had recently published an influential paper on hidden variables with his doctoral advisor Josef-Maria Jauch~\cite{JauchPiron}. Piron objected to the rising fame of Bell's inequality so strongly that Bell's theorem became a ``taboo subject in Piron's circle''~\cite{FreireDissidents}. On his part, Bell criticized the Jauch-Piron paper in his own 1966 analysis of hidden variables~\cite{Bell2}.

The principal reason for this disagreement was the use of subsystems in Bell's theorem. Bell implied a factorizability condition to derive his inequality: in line with EPR, the result of a measurement on a system is unaffected by operations on another, distant system. Measurement results were assigned to subsystems, while relativistic causality was preserved in agreement with Einstein's theory. Piron radically rejected Bell's premise of factorizability. He claimed that if both systems are treated as fully quantum, then taking them as independent subsystems leads to a contradiction. Conceptually, he opposed entanglement and contextuality to Bell-type separability, concluding that quantum systems cannot be partitioned into local subsystems. What was at stake was the very relevance of Einstein's spatiotemporal framework for system independence. Piron's rejection of the  possibility to introduce independent subsystems in quantum theory was an early precursor to the similar tension in the debate on indefinite causality.

\subsection{The definition of agency: a lesson from Hardy}
\label{sect:Hardy}

Following his work on the axioms of quantum theory~\cite{hardy}, Lucien Hardy applied a similar approach to the problem of building an operational theory of quantum gravity. His ``causaloid'' framework~\cite{hardy_probability_2005} is an attempt to reconcile two ``radical features'': 
the irreducibly probabilistic nature of quantum theory with the dynamic causal structure of general relativity. Hardy's solution was operational and axiomatic. Four postulates of the causaloid framework stand out as particularly important~\cite{hardy_probability_2005}:
\begin{itemize}
    \item[1.] \textbf{Inputs and outputs.} A physical theory, whatever else it does, must correlate recorded data. Data consist of (i) a record of actions taken (such as knob settings) and (ii) results of measurements and observations.
    \item[2.] \textbf{Grouping data.} Proximate data are recorded on a card. It will ultimately boil down to a matter of convention and judgement as to what data counts as proximate.
    \item[3.] \textbf{Identity through time.} Each time the experiment is performed, the cards are bundled into a stack and tagged. Since we are interested in constructing a probabilistic theory, we will assume that the experiment can be repeated many times, so that we can construct relative frequencies. The order in which the cards are bundled into any particular stack does not, in itself, represent recorded data.
    \item[4.] \textbf{Infinity jump.} An elementary region, denoted by $R_x$, is the set of all cards taken from the set of \textit{all logically possible} cards, which have some particular $x$ written on them.
\end{itemize}

The first assumption introduces a distinction between inputs and outputs. This postulate lays the foundation for an operational approach, while the fourth assumption, which involves a form of continuity, is characteristic of the axiomatic reconstructions of quantum theory~\cite{grinbaumInfo}. 
The second and third assumptions make data physically meaningful.

The second assumption introduces a grouping of inputs and outputs that belong together, in some sense of proximity. This grouping, called a ``card'' by Hardy, is what we reintepret below as an agent or observer. Unlike their common-language counterparts, these terms do not need to imply human consciousness or any high-level features, like the capacity to will an action. The minimal algebraic view of agency is exactly a collection of inputs and outputs labeled by a name, e.g. Alice or Bob.

The third assumption makes a lawful description possible by introducing repeatability on a set of cards. This structure, called a ``stack'' by Hardy, corresponds to the idea of running the same experiment many times. It might be more intuitive to introduce it, not by composing cards but by dividing the set of all data points that bear the same name (say, $A$ for Alice) into equivalent subgroups. Alice's data points will be indexed by $t_1$, $t_2$, $t_3$, etc., with each index referring to the inputs and outputs in one run. This abstract partition defines physical experiment together with a local time arrow in Alice's laboratory.

Hardy's attempt to build a full-scale physical theory in the causaloid framework  encountered a number of obstacles. Some difficulties appeared due to the need to recover relativistic structures, i.e. a metric, in a fundamentally discrete approach. As he was struggling with these obstacles, Hardy realized that a radical feature of the causaloid framework was its choice of a departure point.
Mathematical construction began with building a global rather than a local Hilbert space. In just a few years, this emphasis on the global Hilbert space was disassociated from searching for a reconstruction of the spacetime metric. The idea has been taken seriously by several researchers who wished to explore a higher-order quantum theory in a global Hilbert space~\cite{oreshkov_quantum_2012,chiribella_quantum_2013}. Hardy's construction of agency, however, remained a foundational step for the operational approach that we carry over in the interpretation of operational eternalism.

\section{Indefinite causality}\label{sect3}

In the standard operational approach, inputs are transformed into outputs in the observer's laboratory via a quantum channel. To connect with the empirical idea of ``reading'' an outcome from a system, a ``physicalization'' axiom is required, i.e. every test can be realized as a channel followed by a measurement on an ancilla system~\cite{Chiribella_2014}. In causality research, inputs and outputs are also distinguished, but time ordering or causal links from the former to the latter are available only locally. While it is possible for Alice to represent her channel as a circuit acting on a system at definite time steps in her local laboratory, it is not possible to extend this intuition globally.

\begin{figure}
    \centering
\includegraphics[width=\linewidth]{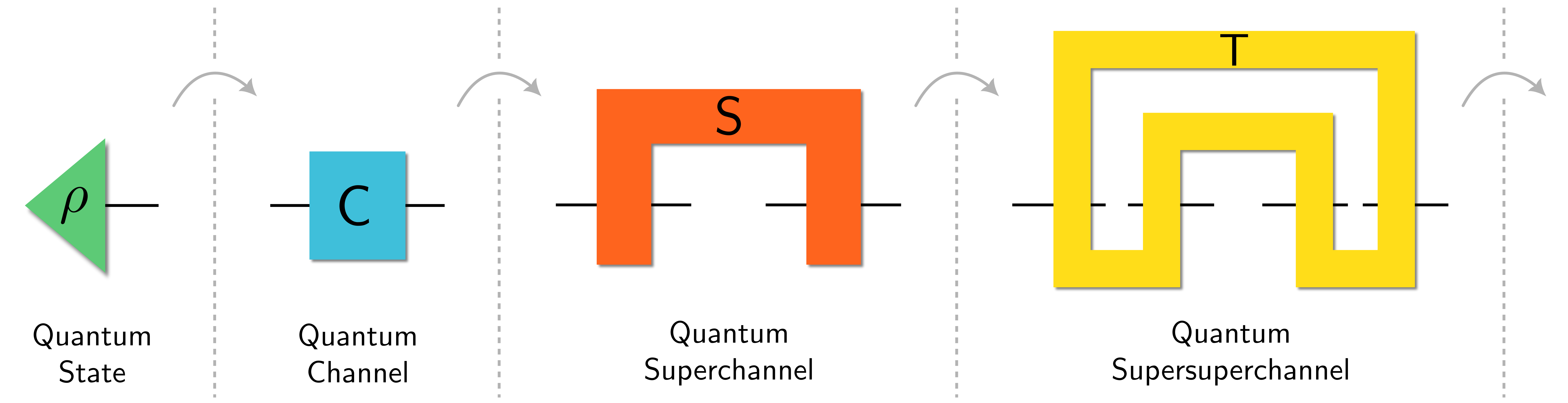}\vfill
\caption{Higher-order processes are generalized processes in a global Hilbert space that includes as subspaces the agents' input and output Hilbert spaces. The agents' channels are transformed by a superchannel in this global space~\cite{taranto2025higherorderquantumoperations}.}
    \label{fig:higher-order}
\end{figure}

\begin{figure}
    \centering
    \includegraphics[height=6.75cm]{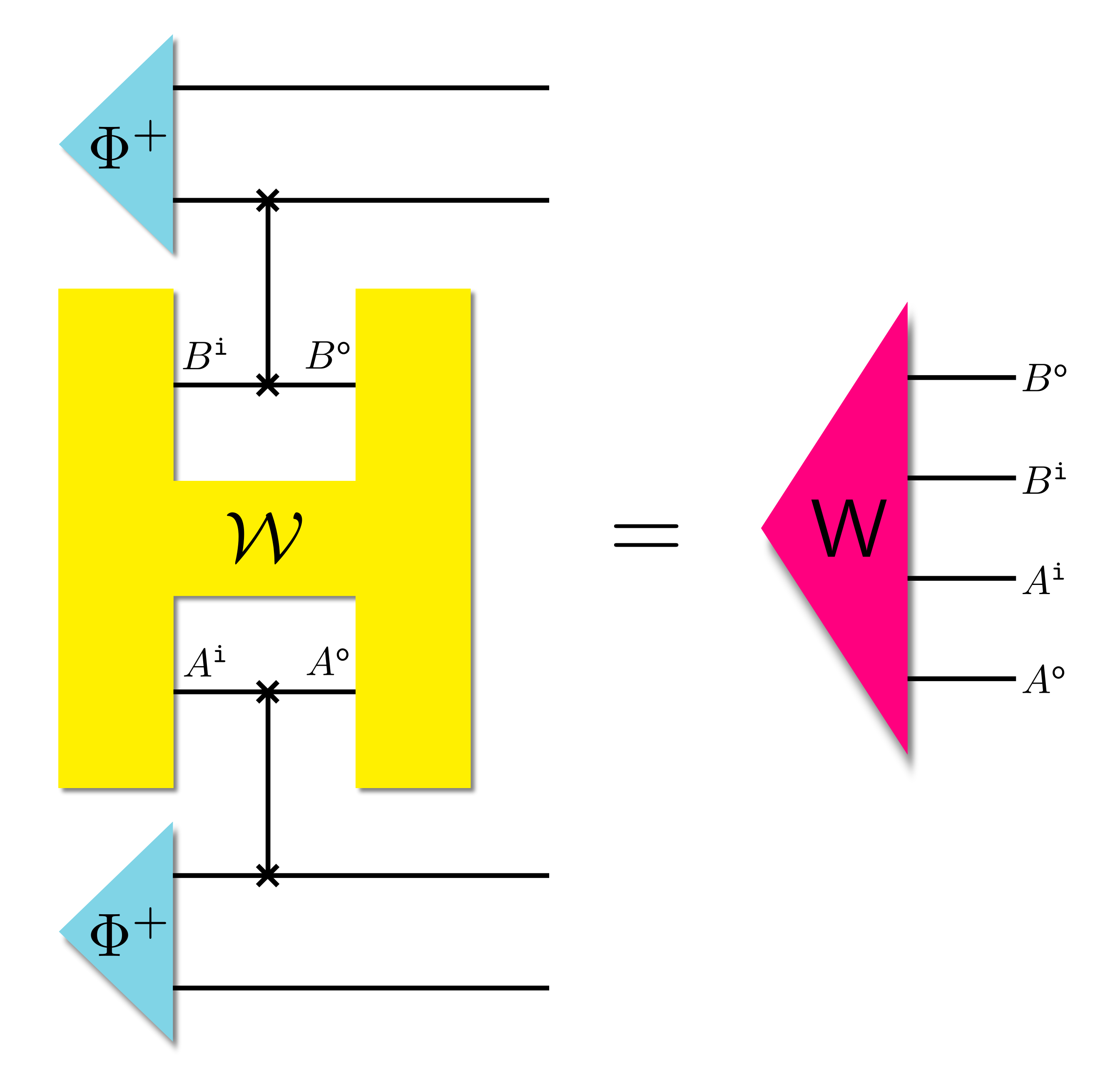}
    \caption{Process matrix~\cite{oreshkov_quantum_2012} in the higher-order process formalism \cite{taranto2025higherorderquantumoperations}. Alice's and Bob's channels relate their respective inputs and outputs, while the higher-order supermap or superchannel $W$ acts on these channels without placing them in any particular time order.}
    \label{fig:processmatrix}
\end{figure}

This is evident from the higher-order picture of processes that transform channels into other channels (Figure~\ref{fig:higher-order})~\cite{taranto2025higherorderquantumoperations}. The global structure of quantum theory is described by a high-dimensional Hilbert space, which is usually (but not necessarily) constructed from the bottom up as a tensor product of the individual agents' input and output Hilbert spaces. This picture yields a theory of supermaps obeying certain consistency constraints~\cite{Chiribella_2008}. Once the global space has been built, a remarkable feature arises: a fully general higher-order theory with indefinite causal order cannot be interpreted as a global sequence of operations performed by local agents on the subspaces that correspond to their inputs and outputs. Rather, supermaps merely connect ``slots'' into which local operations are inserted (Figure~\ref{fig:processmatrix}). They describe physical transformations linking the set of all inputs with the set of all outputs without asking about ``steps'' and without sequentiality. 

The same higher-order construction can also be implemented from the top down by starting with a global Hilbert space without requiring an assignment of inputs and outputs to individual agents. This yields an ``observerless'' or ``observer-neutral''~\cite{wechs2024} perspective that is more general than the tensor-product bottom-up construction. In this perspective, the emergence of agents becomes a matter of theoretical inquiry, as one can consider several agent configurations and choose between them by adding further constraints.

\begin{figure}
    \centering
    \includegraphics[width=6.75cm]{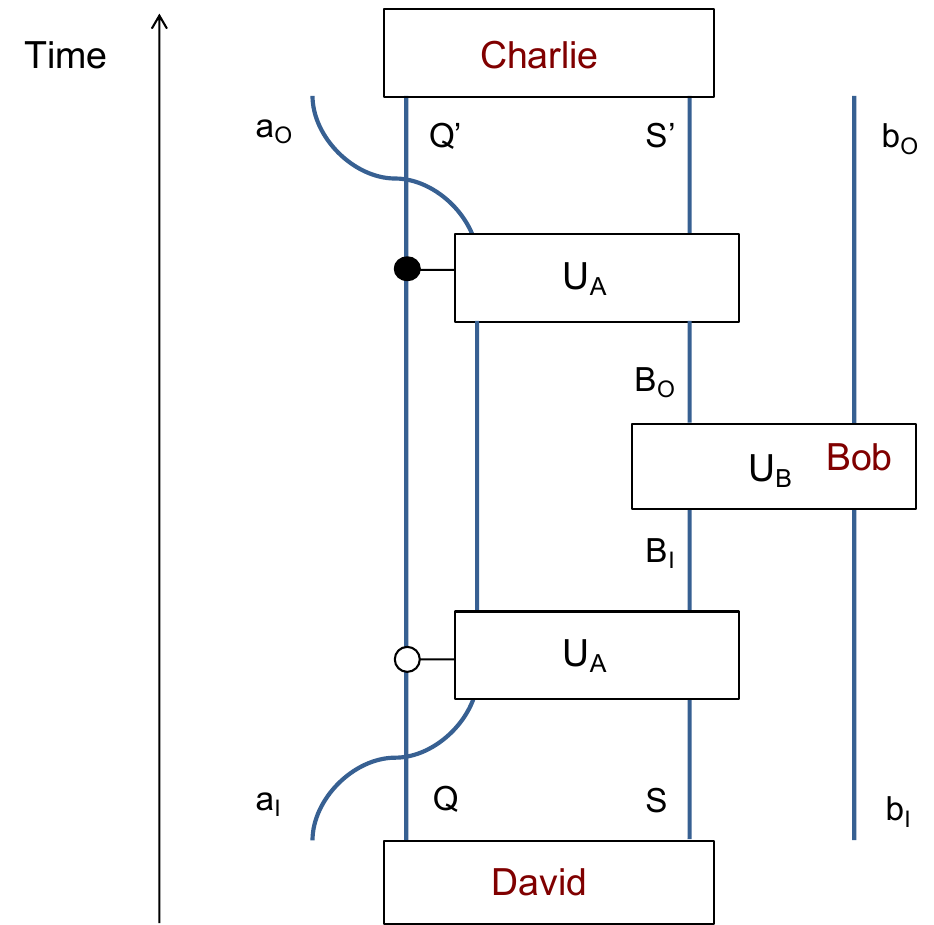}
    \caption{The quantum switch in a circuit representation from Bob's perspective~\cite{Oreshkov2018}. Bob's operation $U_B$ is localized in time, while Alice's input and output subspaces encompass several time steps. Her operation $U_A$ is applied only once but its temporal location is indefinite. David and Charlie are deficient agents added in the common past and future of the setting. They only have an input and an output, respectively, and are equivalent to the observers $P$ and $F$ in the cyclic diagram of the quantum switch in Figure~\ref{fig:cyclic}.}
    \label{fig:delocalized}
\end{figure}

Local time arrows for each agent remain meaningful only in their local frame. No global time unites these perspectives. This lack of a global time arrow precludes a global circuit representation of the process with indefinite causal order. There is no global sequencing of operations performed by agents, and the higher-order operational picture cannot accommodate a vision of observers applying operations in successive time steps. If one tries to enforce a local circuit representation, then it leads to a  process with time-delocalized inputs~\cite{Oreshkov2018,wechs2024}. This pathology is a direct consequence of attempting to force the global structure into a sequential circuit. 
For example, consider the quantum switch in a circuit representation from Bob's perspective (Figure~\ref{fig:delocalized})~\cite{Oreshkov2018}. In this perspective Alice's input and output subsystems encompass several time steps: 
        $$
            \mathcal{H}^{A_I} \subseteq \mathcal{H}^{Q S B_O}\quad 
\mathrm{and}\quad
        \mathcal{H}^{A_O} \subseteq \mathcal{H}^{Q' S' B_I}.$$
They cannot be localized in time by any unitary transformation that respects the sequentiality of Bob's operations according to his local time arrow~\cite{wechs2024}.

Research on indefinite causality in the operational approach has yielded a number of interesting results ranging from computational advantages of indefinite causality~\cite{Bavaresco2021,Brukner2014,Abbott2024,LoizeauAG}, experimental realizations of indefinite causal orders~\cite{Rozema_review,Goswami_review}, foundational aspects of multi-partite settings~\cite{baumeler16,augustin,augustin2}, to possible roads to quantum gravity~\cite{Zych_2019,Castro_Ruiz_2020}. This motivates the need to further explore its conceptual lessons.

Research on indefinite causality has identified a fundamental tension between the view of a global Hilbert space and the sequentiality of locally causal accounts. That the operational primitives---inputs and outputs---no longer map to spacetime events severs the link between the operational approach and spacetime physics. This tension is different from Bell non-locality in time~\cite{LeggettGarg,Brukner_time,Fritz_2010}, as it exemplifies the breakdown of classical causality rather than Leggett-Garg-type entanglement. It is also unrelated to the temporally non-local states that arise in quantum mechanics due to gravity~\cite{Castro_Ruiz_2020} or the uncertainty relations of energy and time~\cite{Filk}. As we argue below, temporal non-locality of the operational primitives hints at the necessity to include a definition of agents within the theory, for it is no longer possible to consider them as metatheoretical entities.

\section{Operational eternalism}\label{sect:eternalism}

\subsection{Analogy with ``block universe''}

Treating all inputs and outputs as jointly embedded in a higher-order supermap bears a structural resemblance to treating all spacetime events on an equal footing in the eternalist, or ``block universe,'' view in the philosophy of spacetime~\cite{tarchi,Sider}. 
Eternalism goes beyond the elimination of a preferred reference frame, which is achieved by relativity alone: it denies objective becoming, removes an arrow of time from fundamental physics, and assigns only a secondary, derivative status to the temporal flow through which events appear to unfold dynamically from past to future. On this view, the primitive constituents of spacetime—namely, events—are considered jointly within a single global perspective.

Similarly, operational eternalism adopts a global perspective in which all inputs and outputs are considered jointly within a single higher-order process. Whereas spacetime eternalism ``freezes geometry,'' operational eternalism ``freezes information.'' The eternalist perspective fixes a global Hilbert space with a process on it, admitting multiple possible assignments of agents without privileging any one of them, just as the block-universe view admits multiple reference frames for a given dynamical process without itself being associated with any particular frame.

The analogy, however, has clear limits. Unlike spacetime eternalism, the interpretation of indefinite causality as operational eternalism carries no ontological commitments: it begins with informational primitives and postulates neither systems nor objects. Instead, it brings agents into the formalism from the metatheoretical role they typically occupy in standard quantum mechanics, making their identification itself an object of theoretical analysis. The higher-order framework thus combines the global perspective suggested by the block universe with an information-theoretic interpretation.

Operational eternalism has three components. First, it treats the global higher-order process as prior to any particular circuit representation. Second, it defines agents through admissible groupings of inputs and outputs. Third, it treats agency as relative to the resulting operational perspectives.

\subsection{Agent from bit}

\begin{figure}
    \centering
        \includegraphics[width=0.325\linewidth]{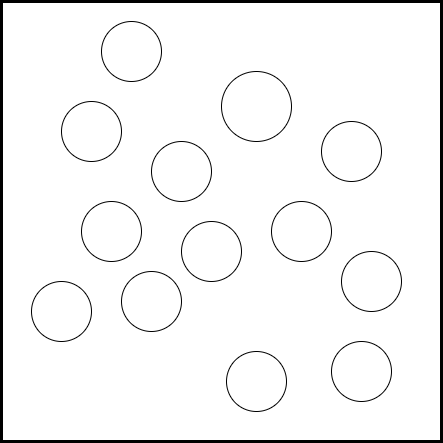}
        \includegraphics[width=0.325\linewidth]{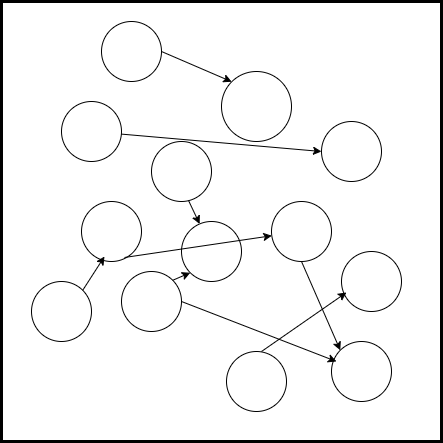}
\includegraphics[width=0.325\linewidth]{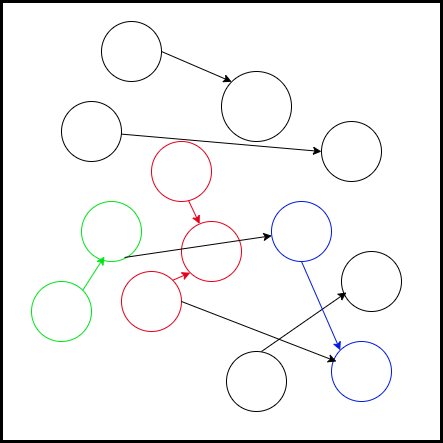}
     \caption{Data under operational eternalism. Left: data points are not initially divided into inputs and outputs. Center: data points are now divided into inputs and outputs, but not yet grouped under observer labels. Right: inputs and outputs are now grouped to distinguish those of Alice (green), Bob (red), Charlie (blue), etc.}
    \label{fig:obs1}
\end{figure}

In operational eternalism, observers or agents are modes of interpreting global data in a consistent way. Inputs and outputs are information-theoretic primitives that do not carry any preassigned observer labels. Assigning both an input and an output to Alice is a constitutive act: this grouping is what defines Alice as an observer.
As Figure~\ref{fig:obs1} illustrates, data points themselves do not belong to anyone; they become ``Alice's input'' or ``Bob's output'' only through a specific grouping. 

Such groupings make theoretical inquiry possible: in a variation on John Wheeler’s famous dictum, the agent arises ``from bit''~\cite{WheelerItFromBit}.
Yet the grouping is not arbitrary. To paraphrase Einstein~\cite{Heis1969}, ``the theory itself should decide'' which groupings or labelings are admissible. Within the higher-order operational framework, a single supermap may admit several such groupings, each yielding a consistent physical interpretation of the same global data~\cite{kushwahaloizeau}. Lawfulness is expressed operationally through the validity conditions imposed on the process matrix. These conditions ensure logical consistency, including for cyclic causal structures, and rule out grandfather-type paradoxes and unrestricted two-way signalling. Whenever these consistency conditions are satisfied, assigning the data points to a set of observers (often called a ``setting'') is operationally meaningful. This does not imply, however, that the resulting ``agents from bit'' can always be assigned geometric reference frames in spacetime. Doing so requires a further step, subject to its own constraints and conditions of possibility, as discussed below.

One could, of course, group the entire global set of inputs and outputs into a single trivial agent and construe the eternalist description as the perspective of a ``godlike'' superobserver. Such a construction, however, would erase causal relations that make up a lawful physical description. The global perspective should therefore be understood not as observerless, but as observer-indefinite. It contains no predefined observers; rather, it contains data that admit multiple groupings into observers. Each admissible grouping yields a different tableau of causal relations.

A further condition on agency concerns the preservation of an observer’s identity throughout an experiment~\cite{bethany2}. As Adlam puts it: 
\begin{quote}
``If a past measurement result could be permanently erased so that no record of it existed in the state of the world at the time of the next measurement, then [\ldots] we could never have all the necessary results available to be compared at the same time''~\cite{Adlam2018}.
\end{quote}
This requirement echoes Everett’s conception of observers as systems endowed with memory, i.e. ``parts... whose states are in correspondence with past experience''~\cite{everett}. Such observers need not be human; they may include ``automatically functioning machines, possessing a sensory apparatus and coupled to recording devices.'' Operationally, preserving Alice’s identity as an observer means preserving her capacity to interpret information as observations of a system. Her recorded inputs and outputs must therefore be identifiable as belonging to one or more runs of the same experiment. In a dynamical description, this requires the relevant memory records to persist under Alice’s time evolution. From a purely operational perspective, however, it is sufficient to specify the appropriate labeling and grouping, i.e. to select suitable subspaces of the global Hilbert space.

At first sight, requiring an agent to possess such a ``history'' appears natural. Yet research on indefinite causality routinely considers observers that cannot record information in memory. Some auxiliary agents, for example, merely prepare a system’s initial state or perform the final measurement~\cite{tein2023,wechs2024,Dourdent2025}.  
These are deficient observers insofar as they lack the memory required to maintain an identity across the experiment. On the account developed here, they should not count as full agents: a single data point—an input or an output merely labeled with Alice’s name—is insufficient to constitute Alice as an agent. Nor is a single data point sufficient to derive a physical law. As suggested by Hardy’s image of a ``stack of cards,'' only a grouping of inputs and outputs permits a lawful theoretical description of the channel connecting them. In a dynamical interpretation, this channel describes the evolution of a physical system within a laboratory.

\subsection{A laboratory in spacetime}

Subject to the consistency and identity conditions described above, any admissible grouping of inputs and outputs can define an observer. Labeling such a grouping with a name such as Alice or Bob constitutes it as a particular agent. Each agent has a local arrow of time and, more generally, a local perspective in which operations admit a circuit representation with inputs preceding outputs. A further question is whether this operational perspective can be geometrized, i.e. represented not merely as a circuit but as a reference frame associated with a laboratory in spacetime. As we shall see, this additional requirement imposes nontrivial constraints on the compatibility of different agents’ perspectives.

Research on indefinite causality sometimes assumes that an agent’s local arrow of time entails the existence of a local laboratory in spacetime, subject to certain ``standard lab assumptions''~\cite{Vilasini_2025}.

One approach associates observers with local regions in a spacetime endowed with a  partial order on events~\cite{vilasini1}. On this account, indefinite causal order arises through coarse-graining an underlying geometry whose causal structure remains definite. A second, more relational approach constructs the laboratory from the ``reference'' degrees of freedom upon which an agent ``is capable, in principle, of making interventions to understand the experiment''~\cite{Vilasini_2025}.
By including operations that are possible in principle, this approach anchors the agent in a space of counterfactual interventions upon a physical object—unperformed operations that could be actualized by the agent. Postulating this space of unperformed operations is a form of entity realism about physical objects. This approach, therefore, retains realist commitments to a geometric conception of agency and to entities that populate the world of physical laboratories, both of which are not contained in the bare higher-order description.

Nevertheless, postulating a laboratory in spacetime does not amount to adopting a na\"ively realist conception of observers. The operational account remains relational with respect to both agents and laboratories. In particular, Bob need not qualify as an agent from within Alice’s local perspective. Bob’s operational primitives, i.e. his inputs and outputs, may be nonlocal for Alice: from her perspective, they are time-delocalized and therefore cannot constitute a coherent agent. No unitary transformation compatible with Alice’s local arrow of time can restore Bob as a well-defined agent within her circuit representation.

Only the global perspective of operational eternalism permits Alice and Bob to be represented simultaneously as agents. Moving from this global description to the local perspective of one agent may deprive an incompatible observer of its agency.
More precisely, in Alice’s causal frame, Bob’s channel is time-delocalized: a single operation on Bob's input or output subspaces appears in a coherent superposition of two temporal slots, one before and one after Alice’s local operation. We denote these two appearances by $Bob_1$ and $Bob_2$; they are not independent agents but two coherently controlled components that become a single localized input or output upon transforming to Bob’s frame. A single unified Bob remains an observer only in his own perspective or other perspectives where his inputs and outputs remain local.

Alice’s representation of $Bob_1$ and $Bob_2$ resembles a Bell-type experiment involving two separate observers who share an entangled system and can communicate in one direction, from $Bob_1$ to $Bob_2$. Yet because this decomposition is defined only relative to Alice’s perspective, the freedom attributed to $Bob_1$ becomes questionable. Unlike an observer in a standard Bell experiment, $Bob_1$
is a deficient perspectival agent whose freedom of choice does not extend across all perspectives. From Bob’s own perspective, by contrast, Bob is a well-defined agent who can, for example, signal to other agents~\cite{augustin2026}.
This perspectival view of agency reveals a mutual incompatibility between operational and spacetime primitives and exposes the limitations of geometry-based formulations of the ``standard lab assumptions.''

A common pool of degrees of freedom may provide a means of reconciling indefinite causality with spacetime, but this strategy does not always succeed. As argued above, observers are fully defined only from the global eternalist perspective, but such  agents need not admit mutually compatible geometrical perspectives. If the operations available in Alice’s and Bob’s laboratories are not localized relative to one another, Alice and Bob will not identify events in the same way. The result is not a single shared spacetime description but several mutually incompatible spacetime perspectives, each consistent with the global eternalist description. Rather than yielding a ``reconciliation between information-theoretic and relativistic views''~\cite{Vilasini_2025}, 
this plurality of incompatible agents brings home a central conceptual lesson of indefinite causality: agency is itself a theoretical construct within the higher-order framework. It is governed by its own metageometrical consistency constraints that are not related to spacetime laboratories. 


\subsection{Friends as mutually compatible agents}

From the standpoint of operational eternalism, an observer becomes fully defined only when inputs and outputs are assigned to a particular grouping. In the absence of a preferred time foliation~\cite{Apadula2026}, no admissible perspective is privileged. Choosing a reference frame or a laboratory selects a particular operational foliation and, with it, a class of compatible co-observers: those groupings whose operations remain time-local relative to that foliation. The same choice may render other groupings time-delocalized, thereby depriving them of the status of coherent observers within the chosen perspective. Compatible observers share mutually consistent arrows of time, can assign compatible temporal labels to their outputs, and can communicate their measurement records classically. Borrowing Wigner’s terminology~\cite{WignerMind}, we call such observers friends.

This suggests the following operational criterion of ``friendliness'': observers $A$ and $B$ are friends if and only if each observer’s operations can be represented as time-localized transformations in the other’s circuit representation. From the global eternalist standpoint, this amounts to requiring a unitary transformation between their perspectives that preserves both operational foliations and hence both local arrows of time. Because the condition is reciprocal, friendliness is symmetric.

\begin{figure}
    \centering
        \includegraphics[width=1\linewidth]{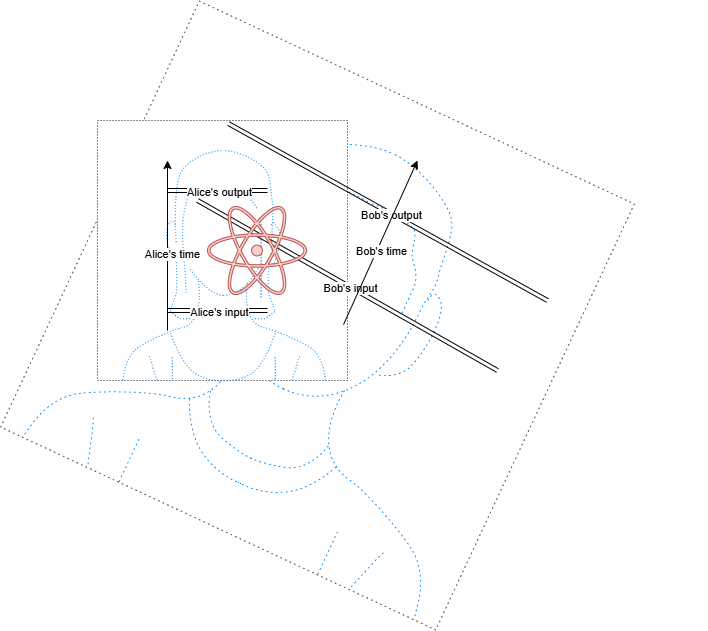}
     \caption{
     In Bob’s laboratory, Bob measures the joint quantum state of Alice and the system, whereas Alice measures only the system. The global eternalist description accommodates both agents, but Alice’s input and output are time-delocalized in Bob’s perspective. Alice and Bob are therefore not friends according to the operational criterion introduced in this section. To exchange their records within a standard Wigner’s-friend scenario, they would need to adopt compatible temporal foliations. However, their perspectives cannot be aligned by a unitary transformation that preserves both local arrows of time.}
        
    \label{fig:wigner}
\end{figure}

The global eternalist description can also accommodate mutually incompatible observers. Such observers are well defined globally, even though no single local perspective represents both of them as time-local agents (Figure~\ref{fig:wigner}). Consequently, they cannot communicate measurement records within a common temporal ordering. In particular, if Alice’s input or output is time-delocalized relative to Bob’s foliation, then Alice and Bob are not friends.

One could, of course, group the entire global set of inputs and outputs into a single trivial agent and construe the eternalist description as the perspective of a ``godlike'' superobserver. Such a construction, however, would erase causal relations that make up a lawful physical description. The global perspective should therefore be understood not as observerless but as observer-indefinite. It contains no predefined observers; rather, it contains data that admit multiple groupings into observers. Each admissible grouping yields a different tableau of causal relations.



\subsection{Agent redefinition may avoid non-causality}

\begin{figure}[h]
    \centering
    \includegraphics[width=0.9\linewidth]{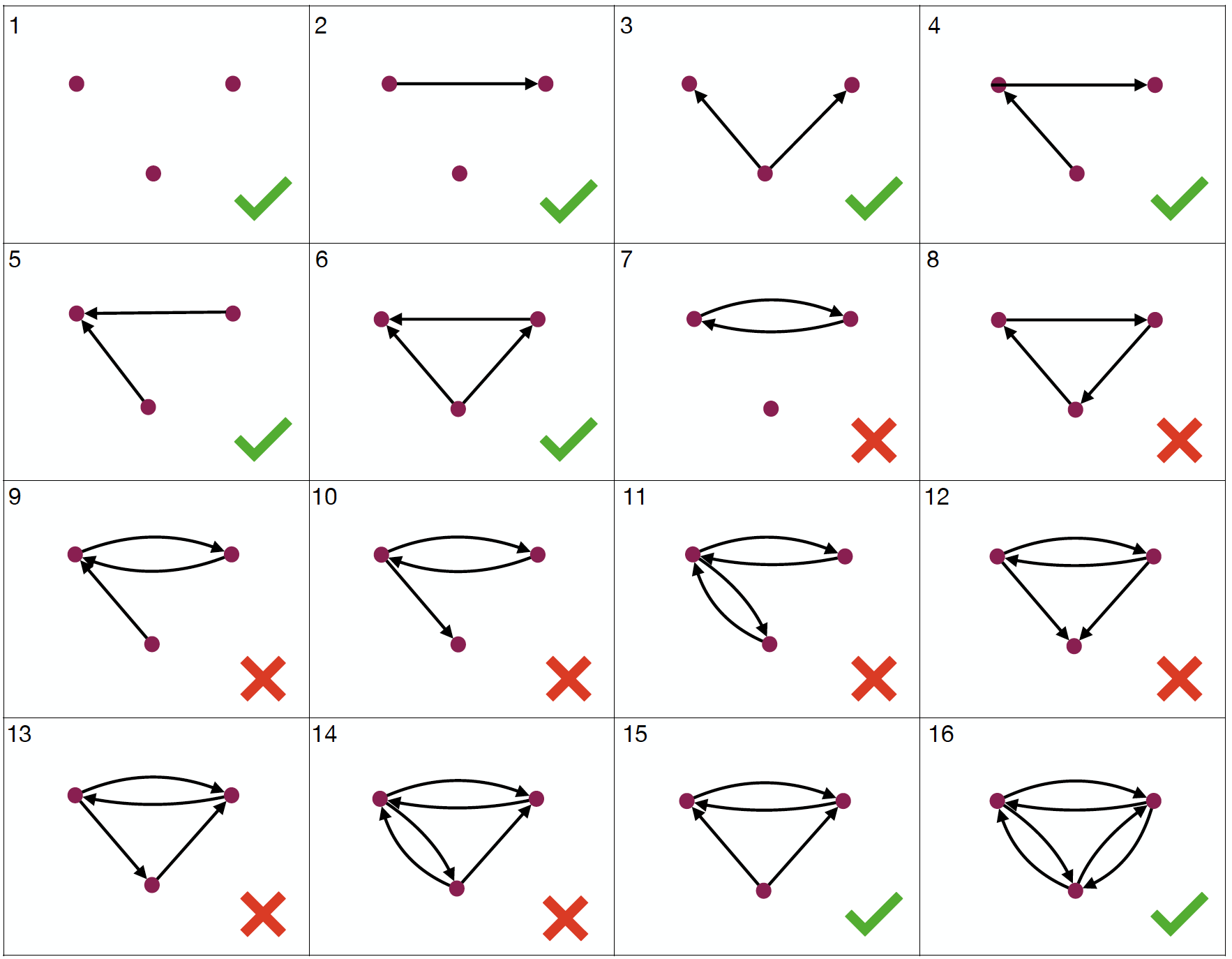}
    \caption{The pairwise non-isomorphic tripartite causal structures classified in~\cite{Tselentis_2023}. Green structures are logically consistent, whereas red structures are inconsistent. Graph 15 represents the quantum switch and graph 16 the Lugano process.}
    
    \label{fig:3partite}
\end{figure}

Logical consistency does not guarantee that a higher-order process can be represented as an ordinary spacetime circuit while preserving a given assignment of inputs and outputs to agents. A process may satisfy all operational consistency conditions and yet admit no circuit realization involving the same agents. The Lugano process—corresponding to one of the eight admissible tripartite structures shown in green in Figure~\ref{fig:3partite}—provides an example. The reason is non-causality: a setting may violate a causal inequality~\cite{oreshkov_quantum_2012,Brukner2015_2,Branciard15,Abbott,tein2023}. 


Like a Bell inequality for non-locality, a causal inequality characterizes correlations compatible with a definite causal order or a probabilistic mixture of definite orders. Its violation shows that the correlations cannot be decomposed into classical causal influences among the agents’ laboratories. The correlations are therefore non-causal relative to the chosen decomposition into agents.
Similarly to the Tsirelson bound for quantum mechanics, quantum higher-order processes violate causal inequalities, not to the algebraic maximum, but up to an intermediate bound that has only been explored computationally~\cite{Liu_2025}.

The first bipartite causal inequality was formulated through the ``Guess Your Neighbor’s Input'' game, together with a higher-order quantum process that violates it~\cite{oreshkov_quantum_2012}.
Known bipartite processes displaying such violations are exceptional: they are not unitarily extendible and hence cannot be obtained, under the usual dilation assumptions, from unitary evolution on a larger Hilbert space~\cite{OreshkovG_2016,Barrett_2021}. The situation changes for three or more agents, where even logically consistent classical processes can violate causal inequalities. In the tripartite Lugano process, for example, each agent is located simultaneously in the past and in the future of every other agent, although the process itself generates no logical contradiction~\cite{baumeler16}.


Such a process remains meaningful within operational eternalism: it is a consistent global assignment of inputs and outputs to several agents. What fails is its representation from the perspective of any one of those agents as an ordinary circuit involving the same set of parties. Any faithful circuit realization must therefore modify the original decomposition into agents.



\begin{figure}[h]
    \centering
    \includegraphics[width=0.25\columnwidth]{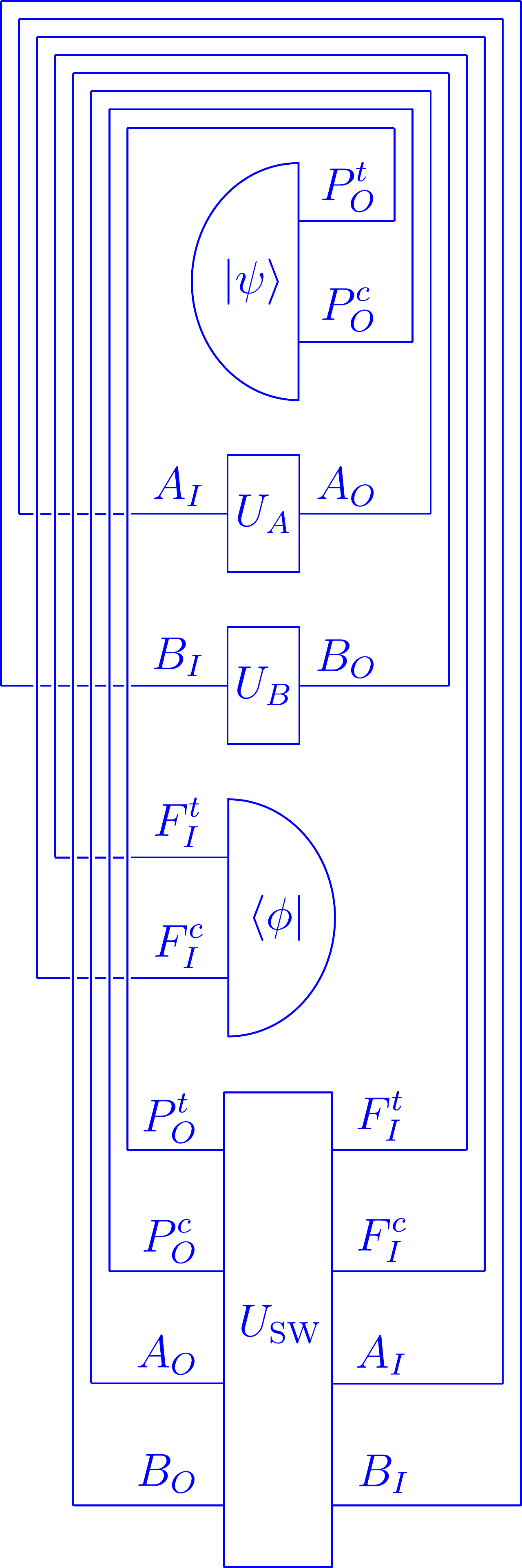}
\caption{A cyclic diagram that describes the quantum switch \cite{wechs2024}. Letters $A$, $B$, $P$ and $F$ label different observers or agents. Systems, including control $c$ and target $t$, propagate along the lines or wires of the diagram. Semi-circular boxes denote preparations or measurements. Rectangular boxes are unitary channels between the agents' inputs ($I$) and outputs ($O$). Note that the global transformation $U_{SW}$ acts on the inputs and outputs of several observers. Deficient observers $P$ and $F$ only have an output and an input, respectively, and sit in the global future and global past of the setting.}
    \label{fig:cyclic}
\end{figure}

This point becomes particularly clear diagrammatically. A cyclic diagram can represent a logically consistent classical or quantum higher-order process even when it contains directed loops (Figure~\ref{fig:cyclic}). A laboratory implementation, by contrast, requires a linear circuit representation relative to some agent’s temporal perspective (Figure~\ref{fig:circuit}). Depending on the process, the required circuit may involve classical or quantum control. Explicit constructions have been given for the Lugano process and the quantum switch~\cite{Dourdent2025,Kunjwal_2023}.


Although the cyclic representation of the Lugano process contains three parties, it cannot be converted into a circuit involving only those same three parties. A circuit realization requires two additional agents, David and Charlie, one in the global past and another in the global future~\cite{Dourdent2025}. David and Charlie play analogous roles in the circuit diagram of the quantum switch (Figure~\ref{fig:delocalized}). The original tripartite description can thus be regarded as a non-causal grouping in a global perspective, which also admits an empirically more realistic five-party setting with a global past and future. The five-party setting no longer violates a causal inequality, although its process may remain causally nonseparable. 
The distinction is analogous to that between entanglement and Bell nonlocality: a process can be causally nonseparable without generating correlations that violate a causal inequality. Unlike in the usual non-local settings, however, indefinite causality allows for enlarging the set of agents, which may trade non-causality for causal non-separability.


Redefining agents is therefore not merely a technical device: it changes the causal interpretation of the process. By regrouping inputs and outputs, one may obtain a decomposition that satisfies causal inequalities~\cite{kushwahaloizeau}, even if this does always make the process causally definite.
The redefinition of agents to avoid a violation of causal inequalities gives precise content to the claim that, in quantum foundations, the observer is the servant of causality. Agents are not metatheoretical entities but non-unique groupings of information. Their theoretical role is to organize global data into operationally coherent structures and, where possible, to provide a lawful sequential account connecting inputs to outputs.

\section{Conclusion}\label{sect:conclusion}

The concept of observer is no longer a metatheoretical assumption as in standard quantum theory. Even in the traditional debate on the interpretation of quantum mechanics, it has gone a long way from the original Copenhagen version possibly related to consciousness~\cite{LB} to the current debate on what constitutes an observer in relational quantum mechanics~\cite{Pienaar_2021,adlam2022}. In indefinite causality research, the observer emerges as a theoretical concept in the sense that it admits a mathematical description with applicable consistency criteria. This concept is metageometrical: it does not involve spatiotemporal considerations. Only the algebraic structure of the theory is at play in the operational approach; any attempt to introduce spacetime primitives, namely pointlike events on a continuous manifold, leads to tension with operational primitives. In a given laboratory, one may not be allowed to consider an observer who is a valid agent in another laboratory. Hence the bare foundation of observerhood belongs with causality, not with geometric methods in physics. 
This signals a paradigm shift: for the first time in the history of quantum theory, the higher-order operational approach makes a mathematical statement about agent assignment.

Operational eternalism is a philosophy that enables the study of agency as groupings of the inputs and outputs. Physical theory begins with a set of data in a higher-order process that does not yet contain the notion of a system or that of an observer. The first task is to group data in consistent ways, often more than one, in order to constitute agents. Until a particular grouping is fixed, agents remain as indefinite as causal relations that are supposed to connect them. This puts the notion of the observer in direct correspondence with avoiding non-causality.

Agents are thus well-defined in the global eternalist picture, but they are not universal across local perspectives. Multiple groupings of data may be admissible. Once a particular agent has been defined, alternative and globally consistent groupings may cease to define agents in their perspective, because respective operational primitives will now appear time-delocalized. This perspectival undoing of agency is a consequence of maintaining the operational idea of sequentiality at the foundation of the circuit paradigm. In any given perspective, one only allows compatible observers that operate in a causally compatible sequence. Compatible observers also provide a criterion of ``friendliness'' in the debate on Wigner's-friend scenarios~\cite{FrauRenner,Cavalcanti2020,Copenhagenish}: friends must share a compatible causal sequence to be able to communicate classically. Operational eternalism suggests that many globally valid observers may not be friends, making it impossible to compare their measurement results.

John Wheeler would have dubbed this approach ``Agent from Bit,'' but it may be possible to go even further down the road of operationalism. For instance, it is possible to imagine that data may not be divided into inputs or outputs. The theory will then tell us which data points correspond to these two categories, as the bipartition will itself become subject to mathematical inquiry~\cite{Chiribella_switching,Stromberg_2024}. Ultimately, the study of physics may not rely on any a priori distinctions or preconditions about spacetime, causality, and agency, not even on historically relativized ones~\cite{Friedman} --- \textit{in the beginning} of the theory \textit{there was data}.

\section*{Acknowledgments}
Many thanks to Luca Apadula, Hippolyte Dourdent, Damiano Santoferrara,  Bethany Terris, and members of the TaQC project for helpful discussions. This research was funded by l’Agence Nationale de la Recherche (ANR), project ANR-22-CE47-0012.

\bibliographystyle{habbrv}

\end{document}